%% file: ANA-HION-2021-17-PAPER.tex
\documentclass[PAPER, cernpreprint, atlasdraft=false,coverpage=false, texlive=2023, UKenglish, texmf, orcidlogo]{atlasdoc}
 
\usepackage{atlaspackage}
\usepackage{atlasbiblatex}
 
\usepackage{atlasphysics}

\graphicspath{{logos/}{figures/}}
 
\usepackage{ANA-HION-2021-17-PAPER-defs}

\AtlasTitle{Strong constraints on jet quenching in centrality-dependent $p$+Pb collisions at $5.02$~TeV from ATLAS}

\AtlasAbstract{
Jet  quenching is the process of color-charged partons losing energy via interactions with quark--gluon plasma droplets created in heavy-ion collisions. The collective expansion of such droplets is well described by viscous hydrodynamics. Similar evidence of collectivity is consistently observed in smaller collision systems, including $pp$ and $p$+Pb collisions.   In contrast, while jet quenching is observed in Pb+Pb collisions, no evidence has been found in these small systems to date, raising fundamental questions about the nature of the system created in these collisions. The ATLAS experiment at the Large Hadron Collider has measured the yield of charged hadrons correlated with reconstructed jets in 0.36~nb$^{-1}$ of $p$+Pb and 3.6~pb$^{-1}$ of $pp$ collisions at 5.02~TeV. The yields of charged hadrons with $p_\mathrm{T}^\mathrm{ch} >0.5$~GeV near and opposite in azimuth to jets with $p_\mathrm{T}^\mathrm{jet} > 30$ or $60$~GeV, and the ratios of these yields between $p$+Pb and $pp$ collisions, $I_{p\mathrm{Pb}}$, are reported. The collision centrality of $p$+Pb events is categorized by the energy deposited by forward neutrons from the struck nucleus. The $I_{p\mathrm{Pb}}$ values are consistent with unity within a few percent for hadrons with $p_\mathrm{T}^\mathrm{ch} >4$~GeV at all centralities. These data provide new, strong constraints which preclude almost any parton energy loss in central $p$+Pb collisions.
}

\AtlasRefCode{HION-2021-17}
 
\PreprintIdNumber{CERN-EP-2022-086}

\AtlasJournalRef{Phys.~Rev.~Lett. 131 (2023) 072301}
\AtlasDOI{10.1103/PhysRevLett.131.072301}

\hypersetup{pdftitle={ATLAS document},pdfauthor={The ATLAS Collaboration}}
 
\begin{document}
 
\maketitle


Over two decades of measurements of relativistic nucleus--nucleus collisions at the Relativistic Heavy Ion Collider (RHIC) and the Large  Hadron Collider (LHC) have established that a quark--gluon plasma is formed in these collisions and undergoes a collective expansion described by viscous hydrodynamics~\cite{Heinz:2013th}.
High-momentum colored probes, such as quarks and gluons, lose some of their energy as they traverse the plasma and produce a highly modified hadron fragmentation pattern, a process referred to as ``jet quenching''~\cite{Cunqueiro:2021wls,Connors:2017ptx,Qin:2015srf}. On the other hand, colorless photons and $Z$ bosons pass through unscathed~\cite{CMS-HIN-19-003,HION-2018-10,ALICE:2020jff,HION-2012-07,CMS-HIN-18-016,HION-2019-05}. Over the past decade, measurements in smaller collision systems,  such as \pp and \pPb at the LHC and $p$+Au, $d$+Au, and $^{3}$He+Au at RHIC display similar experimental evidence of collectivity~\cite{Nagle:2018nvi,Romatschke:2017ejr}. These observations have prompted theoretical discussion as to whether jet quenching should be present in these small systems as well~\cite{Huss:2020dwe,Huss:2020whe,Brewer:2021tyv,Zakharov:2021uza,Zhang:2013oca,Tywoniuk:2014hta,Park:2016jap}. However, measurements of jet~\cite{HION-2013-08,ALICE:2015umm} and hadron~\cite{HION-2012-04,ALICE:2021est,CMS-HIN-15-015} production rates at high transverse momentum (\pT)\footnote{ATLAS uses a right-handed coordinate system with its origin at the nominal interaction point (IP) in the center of the detector and the \(z\)-axis along the beam pipe.   The \(x\)-axis points from the IP to the center of the LHC ring, and the \(y\)-axis points upward.  Cylindrical coordinates \((r,\phi)\) are used in the transverse plane, \(\phi\) being the azimuthal angle around the \(z\)-axis. The pseudorapidity is defined in terms of the polar angle \(\theta\) as \(\eta = -\ln \tan(\theta/2)\).} and measurements of jet-to-hadron fragmentation functions~\cite{HION-2015-04} in minimum-bias \pPb collisions
show no indication of jet quenching, relative to \pp, in these small systems~\cite{Salgado:2016jws}.
 
Experiments have also examined the subset of \pPb collisions where the proton undergoes many interactions in the Pb nucleus (i.e., a large number of proton--nucleon collisions \Ncoll) and which typically have larger-than-average particle multiplicities. These so-called ``central'' events may produce a larger and longer-lived quark--gluon plasma which would induce a bigger jet quenching effect. Measurements which characterize the centrality of events according to the charged-particle multiplicity or energy at mid-rapidity have found significant deviations of high-\pT\ charged-hadron production rates from the \pp expectation~\cite{ALICE:2014xsp}. However, Monte Carlo (MC) simulations using the \textsc{Hijing}~\cite{Gyulassy:1994ew} generator, and other models of small collision systems~\cite{Loizides:2017sqq}, indicate that most of this behavior is the result of physics correlations between the charged-particle multiplicity and the probability to produce a high-\pT jet or hadron in individual proton--nucleon collisions~\cite{ALICE:2014xsp,PHENIX:2013jxf}. In addition, in extreme kinematic regions, such as those with large Bjorken-$x$, the production of high-\pT jets or hadrons becomes anticorrelated with the centrality signal~\cite{HION-2013-08,PHENIX:2015fgy}, which may arise from the decreasing interaction strength of protons in these configurations~\cite{Alvioli:2014eda,Alvioli:2017wou} or from other effects~\cite{Bzdak:2014rca,Kordell:2016njg}. Model-dependent corrections for the effect of these correlations can be derived~\cite{Perepelitsa:2014yta,Loizides:2017sqq}, but they have strongly limited the precision of searches for jet quenching phenomena in central \pPb events. A measurement of the yield of charged hadrons ($8 < \pt < 15$~\GeV) correlated with the highest \pt charged hadron indicate no nuclear modification within uncertainties~\cite{ALICE:2022qxg}.
 
An alternative method, which does not exhibit the physics biases described above, is to select \pPb events by counting the number of spectator neutrons produced by the disintegrating Pb nucleus which strike a zero-degree calorimeter (ZDC), where central events yield more neutrons on average. However, estimating \Ncoll in the resulting event categories is challenging due to limited understanding of how spectator nucleons are distributed in terms of single neutrons, protons, and  larger charged fragments (only the first of which strike the ZDC). Nevertheless, the ALICE Collaboration has used this method and found that rates of charged~\cite{ALICE:2014xsp} and heavy-flavor~\cite{ALICE:2019fhe} hadrons in central \pPb collisions are unmodified in comparison with those derived from \pp interactions, albeit within significant modeling uncertainties. To avoid the reliance on \Ncoll, jet quenching in these events may instead be searched for by examining jet--hadron kinematic correlations or the internal structure of jets. A measurement of hadron-triggered jet yields in ZDC-selected central \pPb events has placed limits on the total amount of energy transported across the boundary of an $R=0.4$ jet cone~\cite{ALICE:2017svf}, which is one possible signature. The measurement presented here can additionally constrain the parton energy loss, even in the case where the jet shape remains unmodified.
 
This Letter presents a measurement of the charged-hadron yield in events with jets in the ATLAS calorimeters satisfying two different selections ($\ptjet > 30$~\GeV\ or $\ptjet > 60$~\GeV) in \pPb and \pp collisions at a $5.02$~\TeV\ nucleon--nucleon center-of-mass energy.
The \pPb and \pp data were recorded in 2016 and 2017, respectively, with triggers sampling integrated luminosities of $\SI{0.36}{nb^{-1}}$ and $\SI{3.6}{pb^{-1}}$. In \pPb running, the proton and lead beams had per-nucleon energies of 4~\TeV\ and $(Z/A)\times4~\TeV \approx 1.58$~\TeV, respectively, leading to a rapidity shift of the center-of-mass frame, $\Delta y^\mathrm{com} = 0.465$, from the laboratory frame (while $y^\mathrm{com} = 0$ in \pp running). Charged hadrons are required to have $\ptch > \SI{0.5}{\GeV}$ and lie within $|\eta-y^\mathrm{com}|<2.035$, and their yields are measured in two azimuthal regions with respect to the jet: the ``away-side'' region $\Delta\phi_\mathrm{ch,jet} = |\phi_{\mathrm{ch}} - \phi_{\mathrm{jet}}| > 7\pi/8$ and the ``near-side'' region $\Delta\phi_\mathrm{ch,jet} < \pi/8$. The total yield in each region, $Y(\ptch)$, is normalized by the number of jets and reported in \pp\ events and in \pPb events for different ZDC energy selections. To quantify any modification which would result from the partons' propagation through a created quark--gluon plasma, the ratio of the per-jet charged-particle yields between \pPb and \pp collisions, $\IpPb = Y_{p\mathrm{Pb}} / Y_{pp}$, is reported and compared with predictions from theoretical calculations. Importantly, this observable does not depend on a quantitative estimate of \Ncoll.

The ATLAS experiment~\cite{PERF-2007-01} is a multipurpose particle detector
with a forward--backward symmetric cylindrical geometry and a near \(4\pi\) coverage in
solid angle.
It consists of an inner tracking detector surrounded by a superconducting solenoid
providing a \SI{2}{\tesla} axial magnetic field, electromagnetic and hadron calorimeters, and a muon spectrometer.
The inner tracking detector covers the pseudorapidity range \(|\eta| < 2.5\).
It consists of silicon pixel, silicon microstrip, and transition radiation tracking detectors.
Lead plus liquid-argon sampling calorimeters provide electromagnetic (EM) energy measurements
with high granularity.
A steel/scintillator-tile hadron calorimeter covers the central pseudorapidity range (\(|\eta| < 1.7\)).
Liquid-argon calorimeters with separate EM and hadronic compartments instrument the endcap (up to $\left|\eta\right| = 3.2$) and forward (FCal, up to $\left|\eta\right| = 4.9$) regions. Two ZDCs are each composed of four longitudinal layers of tungsten absorbers and quartz rods. They are situated in the far forward region \(|\eta| > 8.3\) and, in \pPb events, the downstream ZDC, relative to the Pb beam direction, primarily measures spectator neutrons from the struck Pb nucleus.
An extensive software suite~\cite{ATL-SOFT-PUB-2021-001} is used in the reconstruction and analysis of real and simulated data, in detector operations, and in the trigger and data acquisition systems of the experiment.
 
\begin{figure}  [t]
\centering
\includegraphics[width=0.60\textwidth]{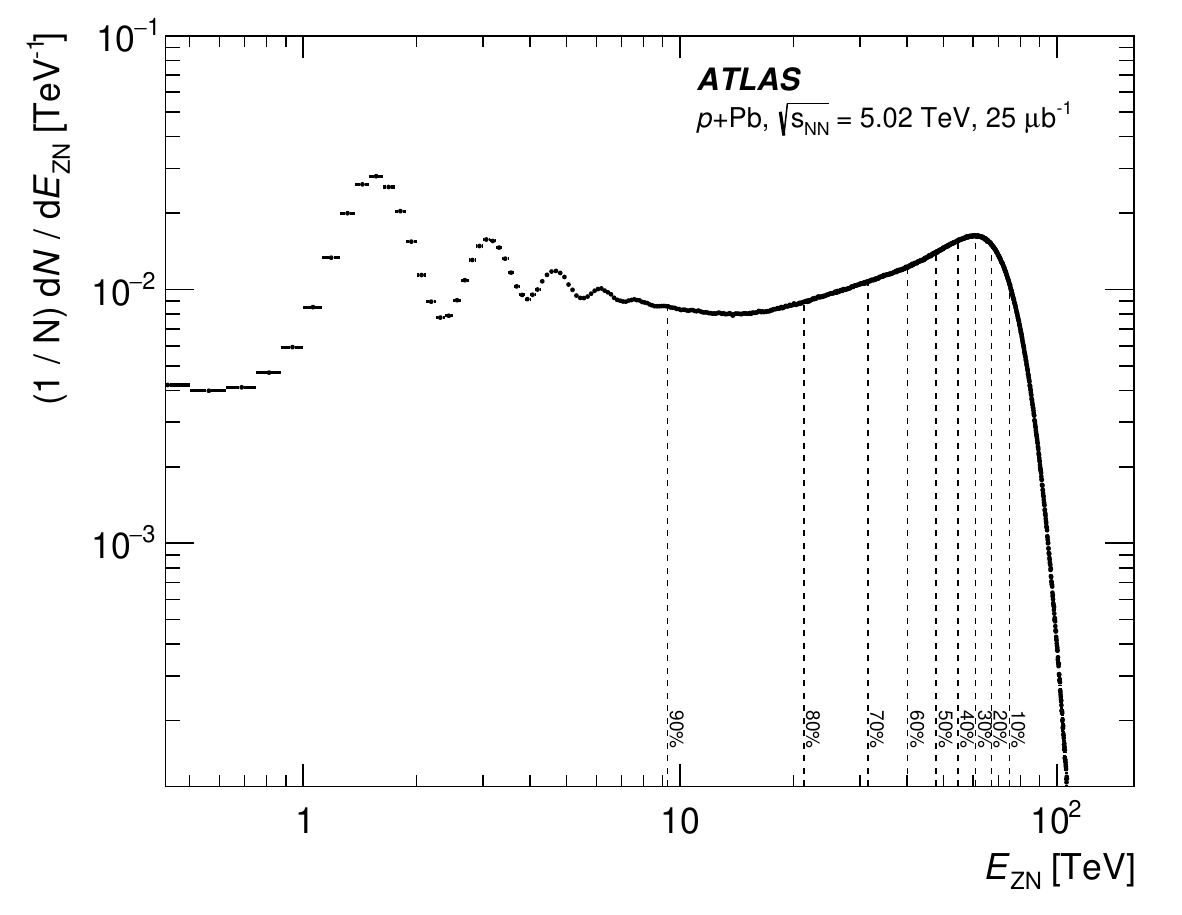}
\caption{
Distribution of energy measured in the Pb-going side of the zero-degree calorimeter ($E_{\mathrm{ZN}}$) in \pPb collisions at 5.02~\TeV\ selected with a minimum-bias trigger. Dashed vertical lines indicate the percentile boundaries between the 0--10\%, 10--20\%, etc., centrality intervals.}
\label{fig:zdc}
\end{figure}
 
Events are selected for analysis using a combination of minimum-bias and calorimeter jet triggers~\cite{TRIG-2016-01}, which are used for the measurements with $\ptjet > 30$~\GeV\ and $>60$~\GeV, respectively. Both the \pp and \pPb events are required to have a primary reconstructed vertex with $z$ coordinate $|z|<150$~mm~\cite{ATL-PHYS-PUB-2015-026}.
The \pp and \pPb data were recorded at low collision rates, and an additional requirement that
events have only one reconstructed interaction vertex further reduces pileup.
In the \pp\ (\pPb) data, this requirement accepts approximately 40\% (99\%) of triggered events.
 
The centrality of \pPb events is characterized using the total energy in the Pb-going side of the ZDC, $E_\mathrm{ZN}$. The ZDC energy is calibrated by matching the single- and double-neutron peaks to their known beam energies ($1.58$~\TeV\ and $3.15$~\TeV)~\cite{HION-2016-02}. The resulting energy distribution is shown in Figure~\ref{fig:zdc}, with more central (lower centrality) events at high $E_\mathrm{ZN}$. Using a ZDC with similar acceptance, the ALICE Collaboration has used multiple bootstrapping methods to estimate that the \Ncoll values in these events range from approximately 13.6 ($\pm11\%$) in 0--20\% centrality \pPb events to 1.2 ($\pm24\%$) in 80--100\% events~\cite{ALICE:2014xsp}, which are likely to be similar in ATLAS. These values and uncertainties are not explicitly used in the measurement in this Letter but may be useful for modeling comparisons.
 
Jets are reconstructed from calorimeter energy deposits as described in Ref.~\cite{HION-2017-06}, using the \antikt\ algorithm~\cite{Cacciari:2008gp,Cacciari:2011ma} with radius parameter $R=0.4$. The jet kinematics are corrected event-by-event for the contribution from underlying event (UE) particles, and are calibrated using simulations~\cite{ATL-SOFT-PUB-2021-001,PERF-2016-04,ATLAS-CONF-2015-016} of the calorimeter response and in situ measurements of the absolute energy scale in data. The accepted jets lie within $|\eta| < 2.8$.
 
Reconstructed charged-particle tracks must satisfy quality criteria outlined in Ref.~\cite{STDM-2015-02}.
The charged-particle yield is corrected for imperfect reconstruction and selection efficiency with a per-track weight, and for the small contribution of secondary-particle and fake tracks, with both corrections derived from \Pythia[8]~\cite{Sjostrand:2014zea} \pp and \textsc{Hijing} \pPb MC event samples.
The contribution of UE particles to the total yields is estimated by measuring these yields in minimum-bias \pPb or \pp events with the same selection requirements and with matched intervals in FCal energy (which is well-correlated with UE activity~\cite{HION-2012-15}). The UE contribution is subtracted from the yield measured in jet-containing events. The ratio of signal to UE background is approximately 0.25 (1) for $\ptch = 0.5$~\GeV\, rising quickly to 3 (30) for $\ptch = 4$~\GeV, in central \pPb (\pp) events. Finally, the finite resolution of the \ptjet and \ptch measurements affects the measured yields. This effect is typically smaller than 10\% and is similar in \pPb and \pp events. It is corrected for via an iterative Bayesian unfolding procedure~\cite{DAgostini:1994fjx} applied to the two-dimensional $(\ptjet, \ptch)$ distributions derived from \Pythia[8] \pp MC events and minimum-bias \pPb data events overlaid with \Pythia[8] \pp events.
 
The dominant sources of systematic uncertainty in the measurement are those affecting the measurement of the jet kinematics, the charged-particle selection, and the unfolding correction. The jet-related uncertainties are derived from in situ studies of the calorimeter response~\cite{PERF-2016-04} and their application to the jets used in heavy-ion data~\cite{ATLAS-CONF-2015-016} (where they accommodate large jet quenching effects), and from comparisons of the simulated response in samples from different generators. They typically dominate at high \ptch. Several sources of tracking-related uncertainty are considered, such as the uncertainty in the absolute efficiency and the sensitivity to selection cuts. They are described in previous measurements of charged-particle fragmentation functions~\cite{HION-2015-04,HION-2017-04} and typically dominate at low \ptch.  The uncertainty in the unfolding correction is evaluated by considering different priors and by performing the analysis procedure, including the UE subtraction, in simulation to evaluate how accurately the generator-level distributions are recovered. This uncertainty is significant at all \ptch. Many of these uncertainties, such as the jet-related ones, have a quantitatively similar impact on the yields in the \pPb and \pp data and largely cancel out in the \IpPb ratio.
 
\begin{figure} [t]
\centering
\begin{tabular}{c}
\includegraphics[width=0.95\textwidth]{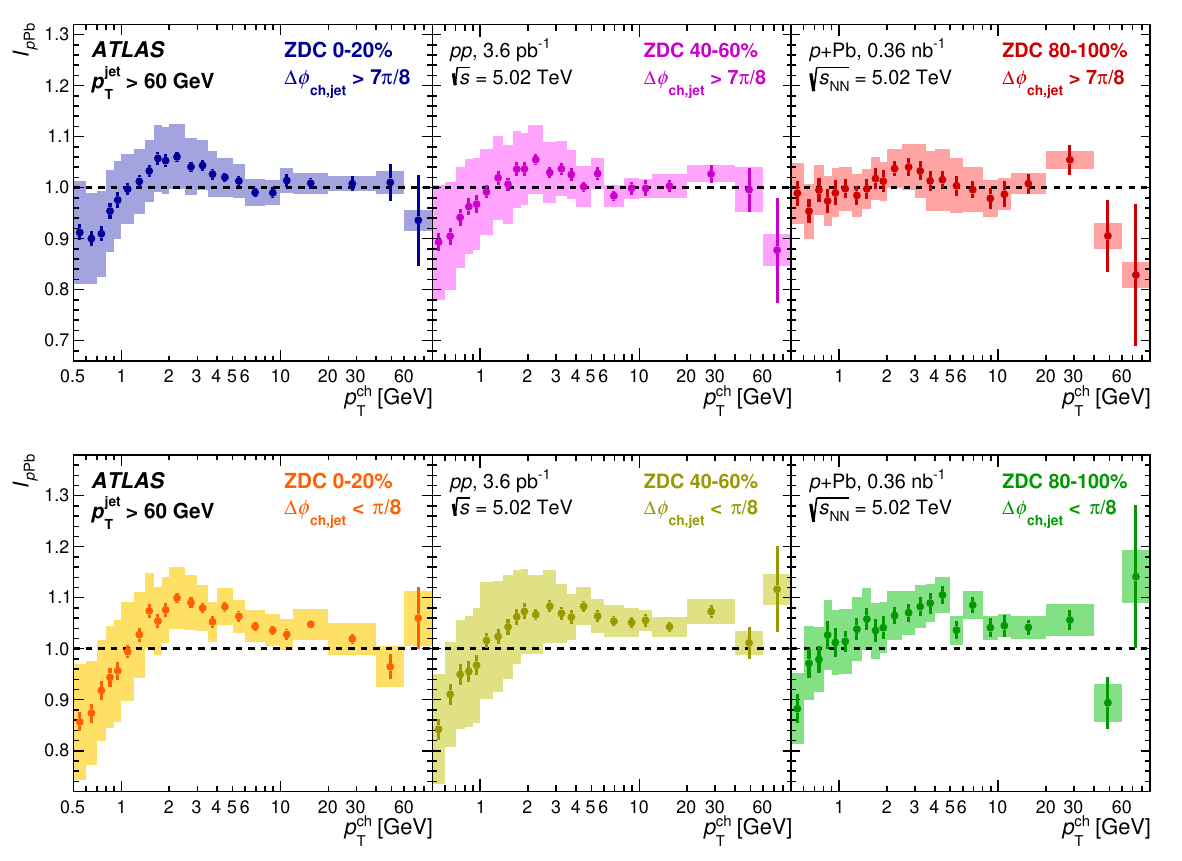}
\end{tabular}
\caption{The ratio of per-jet charged-particle yields between \pPb and \pp collisions, \IpPb, for hadrons opposite ($\Delta\phi_{\mathrm{ch,jet}} > 7\pi/8$; top row) and near ($\Delta\phi_{\mathrm{ch,jet}} < \pi/8$; bottom row) a jet with $\ptjet > 60$~\GeV. Results are shown for different ZDC-selected \pPb centralities in each column. Statistical uncertainties are shown as vertical lines and systematic uncertainties as filled boxes.}
\label{fig:IpPbAll}
\end{figure}
 
Figure~\ref{fig:IpPbAll} (top row) shows $\IpPb$ for charged particles on the away-side of jets with $\ptjet >60$~\GeV. In the region $\ptch > 1$~\GeV, the \IpPb values in all centrality selections are consistent with unity within the uncertainties. At the lowest measured $\ptch$ values, the \IpPb value decreases by about 10\%, albeit with growing uncertainties. In a leading-order parton--parton scattering picture, the away-side hadrons arise from the fragmentation of $\pt \approx 60$~\GeV\ partons azimuthally opposite to the parton producing the jet.
A jet quenching effect in \pPb should lead to \IpPb values below unity.
As such, these results strongly constrain any possible modification of parton fragmentation in the region $z = \ptch/\ptjet \approx 0.05$--$1.0$, within uncertainties that decrease to $2$--$4$\% at high $z$, with respect to that in \pp\ collisions.
 
Figure~\ref{fig:IpPbAll} (bottom row) shows the \IpPb ratios for charged particles on the near-side of jets with \ptjet$>60$~\GeV. In the region \ptch $> 4$~\GeV, there is a centrality-independent enhancement of approximately $5$\%. Similar to that observed on the away side, there is a characteristic suppression in the region $\ptch < 1$~\GeV. Additionally, the \IpPb value shows a modest systematic enhancement in the region $1 < \ptch <4$~\GeV. It is notable that the pattern from 0.5~\GeV\ to 4~\GeV\ is consistent between the away- and near-side, and also with the nuclear modification factor (ratio of total yields between \pPb and \Ncoll-scaled \pp) for inclusive charged hadrons~\cite{HION-2012-14}. The latter is often interpreted in terms of initial-state parton scattering in the nuclear target, also known as the ``Cronin effect''~\cite{PhysRevD.11.3105}.  In heavy-ion collisions, the ``soft'' particle production regime can be described via hydrodynamics and it is known that the radial flow of the quark--gluon plasma may play a role in this \ptch region. The measurement in this Letter suggests that low-\pt particles which arise from jet fragmentation also exhibit a similar pattern.
 
\begin{figure} [ht]
\centering
\begin{tabular}{cc}
\includegraphics[width=0.9\textwidth]{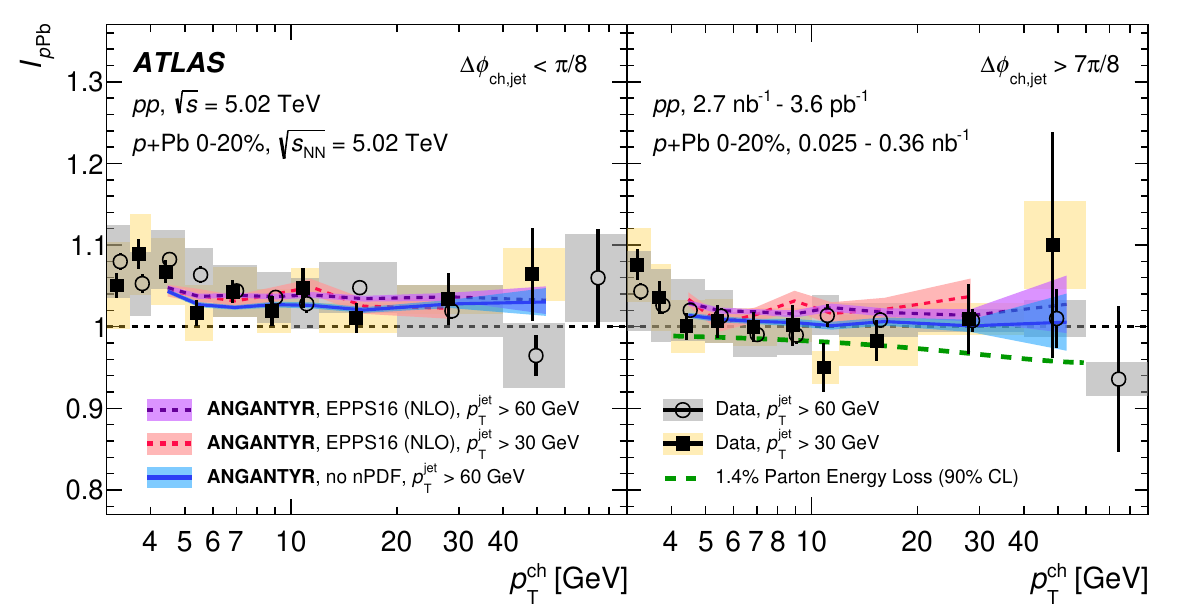}
\end{tabular}
\caption{The ratio of per-jet charged-particle yields, \IpPb, on the near-side (left) and away-side (right) between \pPb and \pp are plotted for the 0--20\% \pPb ZDC-selected centralities. Particles correlated with a jet above 30~\GeV\ and 60~\GeV\ are shown.
Statistical uncertainties are shown as vertical lines and systematic uncertainties as filled boxes.
Also  shown are calculations from the \textsc{Angantyr} generator~\cite{Bierlich:2018xfw} with two treatments of nuclear-modified parton distribution functions (nPDFs): EPPS16~\cite{Eskola:2016oht} at next-to-leading order and no nPDF modification, and with a \Pythia[8]-based model for parton energy loss with vacuum fragmentation (see text).
\textsc{Angantyr} results are only shown for $\ptch > 4.5$~\GeV ~where no UE subtraction is necessary.
The colored bands represent statistical uncertainties only.}
\label{fig:new}
\end{figure}
 
The same measurements were also performed for jets with  \ptjet $> 30$~\GeV, which are sensitive to quenching effects on lower-\pt partons. These are shown in Figure~\ref{fig:new}, focusing on \ptch $> 4$~\GeV\ and 0--20\% centrality ZDC-selected events to emphasize the region of potential jet quenching. Within the larger uncertainties, which arise from the larger relative UE background, poorer jet energy resolution, and smaller sampled luminosity, they are compatible with the $\ptjet > 60$~\GeV\ results. Since the near-side \IpPb is similar to a modified jet fragmentation function, it can be compared with the previous measurement in \pPb collisions by ATLAS~\cite{HION-2015-04}. For jets in a similar \ptjet range, the \pPb-to-\pp ratios of fragmentation functions in Ref.~\cite{HION-2015-04} are compatible with the results in this Letter, although with larger uncertainties due to the different datasets used.
 
In Figure~\ref{fig:new}, the \IpPb measurements are also compared with calculations from the heavy-ion MC generator \textsc{Angantyr}~\cite{Bierlich:2018xfw} run in \pPb mode. \textsc{Angantyr} is based on \Pythia[8] and has no final-state effects producing collectivity or jet quenching -- noting that this is run with so-called ``string shoving'' turned off~\cite{Bierlich:2020naj}.   \textsc{Angantyr} shows a near-side enhancement similar to that in data, and studies with varied generator settings indicate that this does not arise from either the nuclear modification of parton densities or the isospin composition difference between Pb nuclei and protons. On the away-side, the generator features a small enhancement, but is also compatible with the data within its uncertainties.
 
The data is compared to a parton energy loss scenario, modeled using \Pythia[8], where the parton opposite a $\pt > 60$~\GeV\ jet loses a  percentage of its energy before undergoing vacuum-like fragmentation into $\pt > 4$~\GeV\ charged hadrons.   Considering both statistical and systematic data uncertainties, the parton energy loss is constrained to be $0.2 \pm 0.5$\% and less than 1.4\% at the 90\% confidence level (shown in the right panel of Figure~\ref{fig:new}).
 
Despite experimental observations consistent with collectivity in \pPb collisions~\cite{Nagle:2018nvi}, these data severely constrain the amount of jet quenching in central \pPb collisions.
It has been proposed that soft (low-momentum) quarks and gluons are only formed on a time scale of $1$~fm/$c$, and thus the high-\pt partons may undergo their virtuality evolution and showering unscathed and fragment in vacuum if the quark--gluon plasma is small, i.e., with $\textrm{radius} < 1$--$2$~fm~\cite{Muller:2010pm}.    A quantitative calculation incorporating this virtuality evolution is necessary to confront the \IpPb measurements presented here.

In conclusion, this Letter reports a measurement of charged-hadron yields in the azimuthal directions away from and near to jets in \pPb collisions, compared with those in \pp collisions, using data collected with the ATLAS detector at the LHC. Central \pPb collisions, where the effects of a quark--gluon plasma are expected to be largest, are selected in an unbiased way by detecting forward spectator neutrons. The per-jet yields on the near-side indicate a modest, of order 5\%,  enhancement for $\ptch > 4$~\GeV\ that is well described by the MC generator \textsc{Angantyr}. The per-jet yields on the away-side are consistent with unity for all $\ptch > 1$~\GeV, with uncertainties that are particularly small for $\ptch > 4$~\GeV. These data serve as a sensitive probe of jet quenching effects and place strong limits on the degree to which the propagation and fragmentation of hard-scattered partons is modified in small hadronic collisions. The results in this Letter heighten the challenge to the theoretical understanding of the quark--gluon system produced in $p$+Pb collisions.


\section*{Acknowledgments}


We thank CERN for the very successful operation of the LHC, as well as the
support staff from our institutions without whom ATLAS could not be
operated efficiently.
 
We acknowledge the support of
ANPCyT, Argentina;
YerPhI, Armenia;
ARC, Australia;
BMWFW and FWF, Austria;
ANAS, Azerbaijan;
CNPq and FAPESP, Brazil;
NSERC, NRC and CFI, Canada;
CERN;
ANID, Chile;
CAS, MOST and NSFC, China;
Minciencias, Colombia;
MEYS CR, Czech Republic;
DNRF and DNSRC, Denmark;
IN2P3-CNRS and CEA-DRF/IRFU, France;
SRNSFG, Georgia;
BMBF, HGF and MPG, Germany;
GSRI, Greece;
RGC and Hong Kong SAR, China;
ISF and Benoziyo Center, Israel;
INFN, Italy;
MEXT and JSPS, Japan;
CNRST, Morocco;
NWO, Netherlands;
RCN, Norway;
MEiN, Poland;
FCT, Portugal;
MNE/IFA, Romania;
MESTD, Serbia;
MSSR, Slovakia;
ARRS and MIZ\v{S}, Slovenia;
DSI/NRF, South Africa;
MICINN, Spain;
SRC and Wallenberg Foundation, Sweden;
SERI, SNSF and Cantons of Bern and Geneva, Switzerland;
MOST, Taiwan;
TENMAK, T\"urkiye;
STFC, United Kingdom;
DOE and NSF, United States of America.
In addition, individual groups and members have received support from
BCKDF, CANARIE, Compute Canada and CRC, Canada;
PRIMUS 21/SCI/017 and UNCE SCI/013, Czech Republic;
COST, ERC, ERDF, Horizon 2020 and Marie Sk{\l}odowska-Curie Actions, European Union;
Investissements d'Avenir Labex, Investissements d'Avenir Idex and ANR, France;
DFG and AvH Foundation, Germany;
Herakleitos, Thales and Aristeia programmes co-financed by EU-ESF and the Greek NSRF, Greece;
BSF-NSF and MINERVA, Israel;
Norwegian Financial Mechanism 2014-2021, Norway;
NCN and NAWA, Poland;
La Caixa Banking Foundation, CERCA Programme Generalitat de Catalunya and PROMETEO and GenT Programmes Generalitat Valenciana, Spain;
G\"{o}ran Gustafssons Stiftelse, Sweden;
The Royal Society and Leverhulme Trust, United Kingdom.
 
The crucial computing support from all WLCG partners is acknowledged gratefully, in particular from CERN, the ATLAS Tier-1 facilities at TRIUMF (Canada), NDGF (Denmark, Norway, Sweden), CC-IN2P3 (France), KIT/GridKA (Germany), INFN-CNAF (Italy), NL-T1 (Netherlands), PIC (Spain), ASGC (Taiwan), RAL (UK) and BNL (USA), the Tier-2 facilities worldwide and large non-WLCG resource providers. Major contributors of computing resources are listed in Ref.~\cite{ATL-SOFT-PUB-2021-003}.


\clearpage

\printbibliography

\clearpage
 
\input{atlas_authlist}

\end{document}

%% file: atlas_authlist.tex
 
\begin{flushleft}
\hypersetup{urlcolor=black}
{\Large The ATLAS Collaboration}

\bigskip

\AtlasOrcid[0000-0002-6665-4934]{G.~Aad}$^\textrm{\scriptsize 101}$,
\AtlasOrcid[0000-0002-5888-2734]{B.~Abbott}$^\textrm{\scriptsize 119}$,
\AtlasOrcid[0000-0002-7248-3203]{D.C.~Abbott}$^\textrm{\scriptsize 102}$,
\AtlasOrcid[0000-0002-1002-1652]{K.~Abeling}$^\textrm{\scriptsize 55}$,
\AtlasOrcid[0000-0002-8496-9294]{S.H.~Abidi}$^\textrm{\scriptsize 29}$,
\AtlasOrcid[0000-0002-9987-2292]{A.~Aboulhorma}$^\textrm{\scriptsize 35e}$,
\AtlasOrcid[0000-0001-5329-6640]{H.~Abramowicz}$^\textrm{\scriptsize 150}$,
\AtlasOrcid[0000-0002-1599-2896]{H.~Abreu}$^\textrm{\scriptsize 149}$,
\AtlasOrcid[0000-0003-0403-3697]{Y.~Abulaiti}$^\textrm{\scriptsize 116}$,
\AtlasOrcid[0000-0003-0762-7204]{A.C.~Abusleme~Hoffman}$^\textrm{\scriptsize 136a}$
% ...
\bigskip
\\

% ...
$^{29}$Physics Department, Brookhaven National Laboratory, Upton NY; United States of America.\\
% ...
$^{35}$$^{(a)}$Facult\'e des Sciences Ain Chock, R\'eseau Universitaire de Physique des Hautes Energies - Universit\'e Hassan II, Casablanca;$^{(b)}$Facult\'{e} des Sciences, Universit\'{e} Ibn-Tofail, K\'{e}nitra;$^{(c)}$Facult\'e des Sciences Semlalia, Universit\'e Cadi Ayyad, LPHEA-Marrakech;$^{(d)}$LPMR, Facult\'e des Sciences, Universit\'e Mohamed Premier, Oujda;$^{(e)}$Facult\'e des sciences, Universit\'e Mohammed V, Rabat;$^{(f)}$Institute of Applied Physics, Mohammed VI Polytechnic University, Ben Guerir; Morocco.\\
% ...
$^{55}$II. Physikalisches Institut, Georg-August-Universit\"{a}t G\"ottingen, G\"ottingen; Germany.\\
% ...
$^{101}$CPPM, Aix-Marseille Universit\'e, CNRS/IN2P3, Marseille; France.\\
$^{102}$Department of Physics, University of Massachusetts, Amherst MA; United States of America.\\
% ...
$^{116}$Department of Physics, New York University, New York NY; United States of America.\\
% ...
$^{119}$Homer L. Dodge Department of Physics and Astronomy, University of Oklahoma, Norman OK; United States of America.\\
% ...
$^{136}$$^{(a)}$Departamento de F\'isica, Pontificia Universidad Cat\'olica de Chile, Santiago;$^{(b)}$Millennium Institute for Subatomic physics at high energy frontier (SAPHIR), Santiago;$^{(c)}$Instituto de Investigaci\'on Multidisciplinario en Ciencia y Tecnolog\'ia, y Departamento de F\'isica, Universidad de La Serena;$^{(d)}$Universidad Andres Bello, Department of Physics, Santiago;$^{(e)}$Instituto de Alta Investigaci\'on, Universidad de Tarapac\'a, Arica;$^{(f)}$Departamento de F\'isica, Universidad T\'ecnica Federico Santa Mar\'ia, Valpara\'iso; Chile.\\
% ...
$^{149}$Department of Physics, Technion, Israel Institute of Technology, Haifa; Israel.\\
$^{150}$Raymond and Beverly Sackler School of Physics and Astronomy, Tel Aviv University, Tel Aviv; Israel.\\
% ...
$^{a}$ Also Affiliated with an institute covered by a cooperation agreement with CERN.\\
$^{b}$ Also at An-Najah National University, Nablus; Palestine.\\
$^{c}$ Also at Borough of Manhattan Community College, City University of New York, New York NY; United States of America.\\
$^{d}$ Also at Bruno Kessler Foundation, Trento; Italy.\\
$^{e}$ Also at Center for High Energy Physics, Peking University; China.\\
$^{f}$ Also at Centro Studi e Ricerche Enrico Fermi; Italy.\\
% ...

\end{flushleft}
